\documentclass[prb,showpacs,superbib,twocolumn]{revtex4}
\usepackage{amsmath}
\usepackage{graphicx}

\input{tcilatex}

\begin{document}

\title{Conductance modulation in spin field-efect transistors under finite
bias voltages }
\author{Liangbin Hu, Ju Gao, and Shun-Qing Shen}
\affiliation{Department of Physics, The University of Hong Kong, Pokfulam Road, Hong
Kong, China}

\begin{abstract}
The conductance modulations in spin field-effect transistors under finite
bias voltages were studied. It was shown that when a finite bias voltage is
applied between two terminals of a spin field-effect transistor, the spin
precession states of injected spin-polarized electrons in the semiconductor
channel of the device will depend not only the gate-voltage controlled
Rashba spin-orbit coupling but also depend on the bias voltage and, hence,
the conductance modulation in the device due to Rashba spin-orbit coupling
may also depend sensitively on the bias voltage.
\end{abstract}

\maketitle

\section{Introduction}

In the recent years spin-polarized transport in semiconductor
microstructures has attracted much attention because of its important
relevance to the emerging field of spintronics, a new branch of electronics
where the electron's spin ( in addition to its charge ) is the active
element for information storage and processing.\cite{Wolf01} An issue of
fundamental importance in the emerging field of spintronics is the
generation and control of high spin-polarized currents in semiconductors.%
\cite{Wolf01,Prinz98,Kikkawa99,Datta90,Shen03} Recently high efficient
injection of spin-polarized currents from magnetic to non-magnetic
semiconductors have been achieved at low temperatures;\cite{Ohno99} however,
efficient injection of spin-polarized currents from ferromagnetic (F )
metals into semiconductors ( S) has not yet been realized experimentally.
But for room temperature spintronic devices, ferromagnetic metal sources are
indispensable tools. Detailed theoretical investigations have revealed that
the main obstacle for spin injection from a F metal source into a
semiconductor originates from the large mismatch between the conductivities
of metals and semiconductors.\cite{Schmidt00,Johson87} It can be show that
in a usual F metal/semiconductor heterojunction, the spin injection
coefficient is proportional to $\sigma _{S}/\sigma _{F}$, where $\sigma _{S}$
and $\sigma _{F}$ are the conductivity of the semiconductor and the F metal,
respectively. Since $\sigma _{S}<<\sigma _{F}$, the efficiency of spin
injection in usual F metal/semiconductor heterojunctions is very small. At
first glance, this problem seems insurmountable, but very recent theoretical
investigations show that this obstacle may be overcome through the use of
suitable potential barriers\cite{Johson87,Rashba00,Fert01} or through
appropriate epitaxial interfaces that obey certain selection rules and band
structure symmetry properties\cite{Kirczenow01,Mavropoulos02}, and
encouraging experimental results have also been obtained following the
theoretical predictions\cite{Labella01,Zhu01,Hanbicki02}. These results
suggest that devices made of combinations of F metals and semiconductors may
be truly promising for applications in spintronics. Among the most prominent
device proposals that involve combinations of F metals and semiconductors is
the spin field-effect transistor ( spin FET ) \cite{Datta90}. In a spin FET,
two ferromagnetic metallic electrodes are coupled via a ballistic
semiconductor channel. The current modulation in the structure arises from
spin precession of injected spin-polarized electrons in the semiconductor
channel due to Rashba spin-orbit coupling, while two ferromagnetic metallic
electrodes are used to preferentially inject and detect the spin-polarized
currents. It has long been established both theoretically\cite%
{Lommer88,Bychkov84} and experimentally\cite{Luo88,Das89} that, arising from
the structural inversion asymmetry, there is a spin-orbit interaction in two
dimensional electron gases ( 2DEGs ) on narrow-gap semiconductor ( such as
InAs ) surfaces. This underlying spin-orbit interaction was known as Rashba
spin-orbit coupling in the literatures. An important feature of Rashba
spin-orbit coupling is that its strength can by tuned by an external gate
voltage, which alters the build-in structural inversion asymmetry. Due to
this fact, spin precession of injected spin-polarized electrons in the S
channel of a spin FET can be tuned by applying an external gate voltage, and
concomitantly, the current flowing through the device can be also modulated.
This mechanism was first proposed in a seminal work by Datta and Das\cite%
{Datta90} and recently, some important factors that will affect the
behaviors of a spin FET were investigated in more details and with more
realistic assumptions.\cite{Grundler01,Matsuyama01,Larsen02,Molenkamp01,
Mireles01, Schliemann03} In the present paper, we discuss the conductance
modulations in spin FETs under finite bias voltages. Previous theoretical
investigations have been focussed on the zero-bias conductance modulations
in spin FETs, but in practical applications a finite bias voltage need to be
applied between both terminals of a spin FET, and the conductance-bias
voltage characteristics of a device are usually very important for practical
applications of the device. From theoretical viewpoints, when a finite bias
voltage is applied between two terminals of a spin FET, a longitudinal
electric field will be established in the semiconductor channel of the
device, and as was well known, in spin-orbit coupled systems, external
electric field may play a more subtle role on electron's transport than in
traditional electronic devices since in spin-orbit coupled systems the
effect of electric field may be sensitively spin-dependent. ( Examples of
unusual effect of electric field on electron's charge and spin transport in
spin-orbit coupled systems can be seen from Refs.\cite{Hirsch99, Hu03,
Murakami03}. ) In the present paper we discuss the influence of finite bias
voltages on the conductance modulations in spin FETs due to Rashba
spin-orbit coupling. We will show that if a finite bias voltage is applied
between two terminals of a spin FET, the conductance modulation in the
structure due to Rashba spin-orbit coupling may depend sensitively on the
bias voltage, and in order to describe correctly the spin precession state
of injected spin-polarized electrons in the semiconductor channel, the
interplay between the Rashba spin-orbit coupling ( which can be tuned via
the gate voltage ) and the longitudinal electric field induced by the
application of a finite bias voltage should be described in a unified way.

\section{Model and Formulation}

For simplicity, in this paper we will restrict ourselves to a
one-dimensional (1D) model. In one-dimensional systems the quantum
interference effect due to Rashba spin-orbit coupling will be maximum since
the phase shifts of electrons are independent of their paths, so the
idealized 1D model will give an upper limit for the achievable
spin-transistor effect. In higher dimensions, the phase shifts of electrons
will depend on their paths and, hence, the spin-transistor effect will
become weaker than what is predicted in a 1D model system. This was
illustrated in Ref.\cite{Hu01}. Though in the present paper we restrict our
discussion to a 1D model system, the formulas given below are easy to be
extended to systems with higher dimensions. This will be discussed
elsewhere. In the one-band effective-mass approximation, the 1D model system
can be described by the following Hamiltonian: 
\begin{align}
\hat{H}& =\frac{1}{2}\hat{p}_{x}\frac{1}{m(x)}\hat{p}_{x}+\frac{1}{2\hbar }%
\hat{\sigma}_{z}[\hat{p}_{x}\alpha (x)+\alpha (x)\hat{p}_{x}]  \notag \\
& +\frac{1}{2}\Delta \mathbf{\hat{\sigma}\cdot }[\vec{m}_{L}\theta (-x)+\vec{%
m}_{R}\theta (x-L)]+\delta E_{c}\theta (x)\theta (L-x)  \notag \\
& +\hat{U}[\delta (x)\mathbf{+}\delta (x-L)]\mathbf{+}V(x).
\end{align}%
Here $\theta (x)$ is the usual step function and $\delta (x)$ the usual $%
\delta $ function, $\hat{p}_{x}$ is the momentum operator, $\mathbf{\hat{%
\sigma}}$ are the Pauli matrices, $m(x)=m_{f}+(m_{s}-m_{f})$ $\theta
(x)\theta (L-x)$ is the effective mass of electron, with $m_{f}$ denoting
the effective mass of electron in the ferromagnetic contacts and $m_{s}$ the
effective mass of electron in the semiconductor channel, and the F/S
interfaces are assumed to be located at $x=0$ and $x=L$. The second term in
Eq.(1) describes the Rashba spin-orbit coupling\cite%
{Matsuyama01,Larsen02,Molenkamp01, Mireles01}, where $\alpha (x)$ is defined
by $\alpha (x)\equiv \alpha _{R}\theta (x)\theta (L-x)$, and $\alpha _{R}$
is the Rashba spin-orbit coupling constant in the S region. Since the
Hamiltonian $\hat{H}$ has to be an Hermitian operator, in Eq.(1) we have
used the symmetrized version of Rashba spin-orbit interaction. The third
term in Eq.(1) describes the exchange interaction in the ferromagnetic
contacts, with $\Delta $ denoting the spin-splitting energy and the unit
vector $\vec{m}_{L}$ ( $\vec{m}_{R}$ ) denoting the direction of the
magnetization in the left ( right) contact. It will be assumed that $\vec{m}%
_{L}$ is in the $+x$ direction and $\vec{m}_{R}$ will be in either $+x$
direction ( parallel configuration ) or $-x$ direction ( antiparallel
configuration). The fourth and fifth terms in Eq.(1) model the conduction
band mismatch and the interfacial scattering between the F and S regions,
respectively, with $\delta E_{c}$ denoting the band mismatch and $\hat{U}$
the interfacial scattering potential. In the presence of both
spin-conserving and spin-flip interfacial scattering, $\hat{U}$ will be a $%
2\times 2$ matrix with the diagonal elements $(U^{\uparrow \uparrow
},U^{\downarrow \downarrow })$ representing the spin-dependent strength of
spin-conserving interfacial scattering and the off-diagonal elements $%
(U^{\uparrow \downarrow },U^{\downarrow \uparrow })$ the strength of
spin-flip interfacial scattering. For simplicity, we will assume that $%
U^{\uparrow \uparrow }=U^{\downarrow \downarrow }$ and $U^{\uparrow
\downarrow }=U^{\downarrow \uparrow }$. ( For magnetically active interface,
its is possible that $U^{\uparrow \uparrow }\neq U^{\downarrow \downarrow }$
and $U^{\uparrow \downarrow }\neq U^{\downarrow \uparrow }$. ) Finally, the
last term in Eq.(1) denotes the longitudinal electric potential due to the
application of a finite bias voltage, and the longitudinal electric
potential is given by $V(x)=-eV_{0}\theta (x-L)-eV_{0}(x/L)\theta (x)\theta
(L-x)$, where $V_{0}$ is the magnitude of the applied bias voltage. Due to
the application of the bias voltage $V_{0}$, a longitudinal electric field $%
F\equiv V_{0}/L$ will be established in the semiconductor channel of the
structure and the Fermi energy $\mu _{R}$ in the right contact will be
lowered by e$V_{0}$ with respect to the Fermi energy $\mu _{L}$ in the left
contact.

To obtain the spin conductance of the device described by the Hamiltonian
(1), we start by considering the scattering problem related to the
interfaces between the F and S regions. In order to solve the scattering
problem, one need to find first the eigenstates in each region. In the
ferromagnetic contacts ( $x<0$ and $x>L$ ), one obtains from the Hamiltonian
(1) the eigenstates with energy $E$, 
\begin{eqnarray}
\Psi _{F,\sigma ,L}^{(\pm )} &=&\phi _{F,\sigma ,L}^{(\pm )}(x)|\sigma
\rangle ,\,\phi _{F,\sigma ,L}^{(\pm )}(x)  \notag \\
&=&\sqrt{\frac{m_{f}}{\hbar k_{\sigma ,L}}}e^{\pm ik_{\sigma ,L}x},\;(x<0),
\\
\Psi _{F,\gamma ,R}^{(\pm )} &=&\phi _{F,\gamma ,R}^{(\pm )}(x)|\gamma
\rangle ,\,\phi _{F,\gamma ,R}^{(\pm )}(x)  \notag \\
&=&\sqrt{\frac{m_{f}}{\hbar k_{\gamma ,R}}}e^{\pm ik_{\gamma ,R}x},\;(x>L),
\end{eqnarray}%
where $|\sigma \rangle $ ( $\sigma =\pm $ ) and $|\gamma \rangle $ ( $\gamma
=\pm $ ) are the spinor eigenstates in the left and right ferromagnetic
contacts, respectively, which are defined by 
\begin{subequations}
\begin{eqnarray}
\{|+\rangle _{L},|-\rangle _{L}\} &=&\frac{1}{\sqrt{2}}\left( 
\begin{array}{l}
\pm 1 \\ 
1%
\end{array}%
\right) , \\
\{|+\rangle _{R},|-\rangle _{R}\} &=&\lambda \frac{1}{\sqrt{2}}\left( 
\begin{array}{l}
\pm 1 \\ 
1%
\end{array}%
\right) ,
\end{eqnarray}%
where $\lambda =+1$ if the two ferromagnetic electrodes are in parallel
configuration and $\lambda =-1$ if the two electrodes are in antiparallel
configuration. The wave number $k_{\sigma ,L}$ ( $k_{\gamma ,R}$ ) will be
given by $k_{\pm ,L(R)}=\sqrt{\frac{2m_{f}}{\hbar ^{2}}(E\mp \Delta )}$ .
The eigenfunctions in the S region cannot be written down directly from the
Hamiltonian (1) due to the presence of the last term in Eq.(1). To find the
eigenstates in the S region, we first note that in the S region the
Hamiltonian (1) is spin-diagonal and the eigenstates have the form $\Psi
_{S,\beta }(x)=\phi _{S,\beta }(x)|\beta \rangle $ and $\Psi _{S,\bar{\beta}%
}(x)=\phi _{S,\bar{\beta}}(x)|\bar{\beta}\rangle $ , where $|\beta \rangle
=(1,0)$ and $|\bar{\beta}\rangle =(0,1)$ are the spinor eigenstates in the S
region. The Schr\"{o}dinger equation in the S region will reduce to 
\end{subequations}
\begin{eqnarray}
&&-\frac{\hbar ^{2}}{2m_{s}}\frac{\partial ^{2}}{\partial x^{2}}\phi
_{S,\beta }(x)-i\alpha _{R}\frac{\partial }{\partial x}\phi _{S,\beta }(x)-%
\frac{eV_{0}x}{L}\phi _{S,\beta }(x)  \notag \\
&=&E\phi _{S,\beta }(x), \\
&&-\frac{\hbar ^{2}}{2m_{s}}\frac{\partial ^{2}}{\partial x^{2}}\phi _{S,%
\bar{\beta}}(x)+i\alpha _{R}\frac{\partial }{\partial x}\phi _{S,\bar{\beta}%
}(x)-\frac{eV_{0}x}{L}\phi _{S,\bar{\beta}}(x)  \notag \\
&=&E\phi _{S,\bar{\beta}}(x).
\end{eqnarray}%
After making a transformation $\phi _{S,\beta }(x)\rightarrow w_{\beta
}(x)=\phi _{S,\beta }(x)e^{i\alpha _{R}mx/\hbar ^{2}}$ and $\phi _{S,\bar{%
\beta}}(x)\rightarrow w_{\bar{\beta}}(x)=\phi _{S,\bar{\beta}}(x)e^{-i\alpha
_{R}mx/\hbar ^{2}}$, it can be shown that both $w_{\beta }(x)$ and $w_{\bar{%
\beta}}(x)$ will satisfy the following equation

\begin{equation}
\frac{\partial ^2}{\partial x^2}w(x)+\frac{2eV_0m_s}{L\hbar ^2}(x+\epsilon
_0)w(x)=0,
\end{equation}
where $\epsilon _0$ is defined by 
\begin{equation}
\epsilon _0=\frac{EL}{eV_0}+\frac{\alpha _R^2m_sL}{2eV_0\hbar ^2}.
\end{equation}
Eq.(7) can solved with the help of the Airy functions and the two linearly
independent solutions can be given by $Ai[-(2eV_0m_s/L\hbar
^2)^{1/3}(x+\epsilon _0)]$ and $Bi[-(2eV_0m_s/L\hbar ^2)^{1/3}(x+\epsilon
_0)]$. Here $Ai[z]$ and $Bi[z]$ are the usual Airy functions\cite{Abramowitz}%
. Then one can see that there are four eigenstates in the S region, and the
corresponding eigenfunctions $\Psi _{S,\beta }^{(i)}(x)$ and $\Psi _{S,\bar{%
\beta}}^{(i)}(x)$ ( $i=1,2$ ) will be given by 
\begin{eqnarray}
\Psi _{S,\beta }^{(i)}(x) &=&\phi _{S,\beta }^{(i)}(x)|\beta \rangle ,\,\phi
_{S,\beta }^{(i)}(x)=e^{-i\alpha _Rmx/\hbar ^2}w^{(i)}(x),\, \\
\Psi _{S,\bar{\beta}}^{(i)}(x) &=&\phi _{S,\bar{\beta}}^{(i)}(x)|\bar{\beta}%
\rangle ,\phi _{S,\bar{\beta}}^{(i)}(x)=e^{i\alpha _Rmx/\hbar ^2}w^{(i)}(x),
\end{eqnarray}
where $w^{(1)}(x)\equiv Ai[-(2eV_0m_s/L\hbar ^2)^{1/3}(x+\epsilon _0)]$ and $%
w^{(2)}(x)\equiv Bi[-(2eV_0m_s/L\hbar ^2)^{1/3}(x+\epsilon _0)]$.

Now we consider the scattering state of an electron with energy $E$ and spin 
$\sigma $ incoming from the ferromagnetic lead ( $x<0$ ). The total wave
function including the reflected and transmitted waves can be written as: 
\begin{eqnarray}
\Psi _{F}(x) &=&\phi _{F,\sigma ,L}^{(+)}(x)|\sigma \rangle +r_{\sigma
\sigma }\phi _{F,\sigma ,L}^{(-)}(x)|\sigma \rangle   \notag \\
&+&r_{\sigma \bar{\sigma}}\phi _{F,\bar{\sigma},L}^{(-)}(x)|\bar{\sigma}%
\rangle ,\quad x<0, \\
\Psi _{S}(x) &=&\sum_{i=1,2}c_{i,\beta }\phi _{S,\beta }^{(i)}(x)|\beta
\rangle   \notag \\
&+&\sum_{i=1,2}c_{i,\bar{\beta}}\phi _{S,\bar{\beta}}^{(i)}(x)|\bar{\beta}%
\rangle ,\,0<x<L, \\
\Psi _{F}(x) &=&t_{\sigma \gamma }\phi _{F,\gamma ,R}^{(+)}(x)|\gamma
\rangle   \notag \\
&+&t_{\sigma \bar{\gamma}}\phi _{F,\bar{\gamma},R}^{(+)}(x)|\bar{\gamma}%
\rangle ,\quad x>L,
\end{eqnarray}%
where $r_{\sigma \sigma }$, $r_{\sigma \bar{\sigma}}$, $t_{\sigma \gamma }$, 
$t_{\sigma \bar{\gamma}}$, $c_{i,\beta }$, and $c_{i,\bar{\beta}}$ ( $i=1,2$
) are coefficients that need to be determined by the boundary conditions.
The matching conditions at the interfaces between the F and S regions can be
obtained by integrating $\hat{H}\Psi =E\Psi $ from $x=-\varepsilon $ to $%
x=+\varepsilon $ and from $x=L-\varepsilon $ to $x=L+\varepsilon $ in the
limit $\varepsilon \rightarrow 0$. This yields 
\begin{align}
\Psi _{F}(x)|_{x=0^{-}}& =\Psi _{S}(x)|_{x=0^{+}}, \\
\Psi _{S}(x)|_{x=L^{-}}& =\Psi _{F}(x)|_{x=L^{+}}, \\
\hat{v}_{S}\Psi _{S}(x)|_{x=0^{+}}& =\hat{v}_{F}\Psi _{F}(x)|_{x=0^{-}}-%
\frac{2i}{\hbar }\hat{U}\Psi _{F}(x)|_{x=0^{-}}, \\
\hat{v}_{S}\Psi _{S}(x)|_{x=L^{-}}& =\hat{v}_{F}\Psi _{F}(x)|_{x=L^{+}}+%
\frac{2i}{\hbar }\hat{U}\Psi _{F}(x)|_{x=L^{+}},
\end{align}%
where $\hat{v}_{F}=\hat{p}_{x}/m_{f}$ and $\hat{v}_{S}=\hat{p}%
_{x}/m_{s}+(\alpha _{R}/\hbar )\hat{\sigma}_{z}$ are the velocity operators
in the F and S regions, respectively. From the matching conditions
(14)-(17), the transmission coefficients $t_{\sigma \gamma }$ can be
obtained. Then in the linear response regime and in the low temperature
limit, the spin conductance $G_{\sigma }$ and the total conductance $G$ of
the device can be calculated through the Landauer formula, given by 
\begin{equation}
G=\sum_{\sigma =\pm }G_{\sigma },\,G_{\sigma }=\frac{e^{2}}{h}\sum_{\gamma
=\pm }|t_{\sigma \gamma }(\mu )|^{2},\,
\end{equation}%
where $\mu $ is the average of the Fermi energies $\mu _{L}$ and $\mu _{R}$
on the left and right electrodes, respectively\cite{Ando98}. The spin
injection efficiency for the device can be defined by $\eta
=(G_{+}-G_{-})/(G_{+}+G_{+})$. This ratio characterizes the spin
polarization of the electric current flowing through the device. The
conductance of the device and the spin injection coefficient will depend on
the magnetization configurations in the two ferromagnetic electrodes. In the
following we will denote the conductance as $G^{(P)}$ and the spin injection
coefficient as $\eta ^{(P)}$ if the magnetizations in the two electrodes are
parallel and as $G^{(AP)}$ and $\eta ^{(AP)}$ if the magnetizations in the
two electrodes are antiparallel. The change in conductance when the two
ferromagnetic electrodes switch between parallel and antiparallel
configurations can be measured by a magnetoconductance ratio $\eta _{M}$,
defined by 
\begin{equation}
\eta _{M}=\frac{G^{(P)}-G^{(AP)}}{G^{(P)}+G^{(AP)}}.
\end{equation}

\section{Results and Discussions}

\begin{figure}[tbp]
\includegraphics*[width=8.7cm]{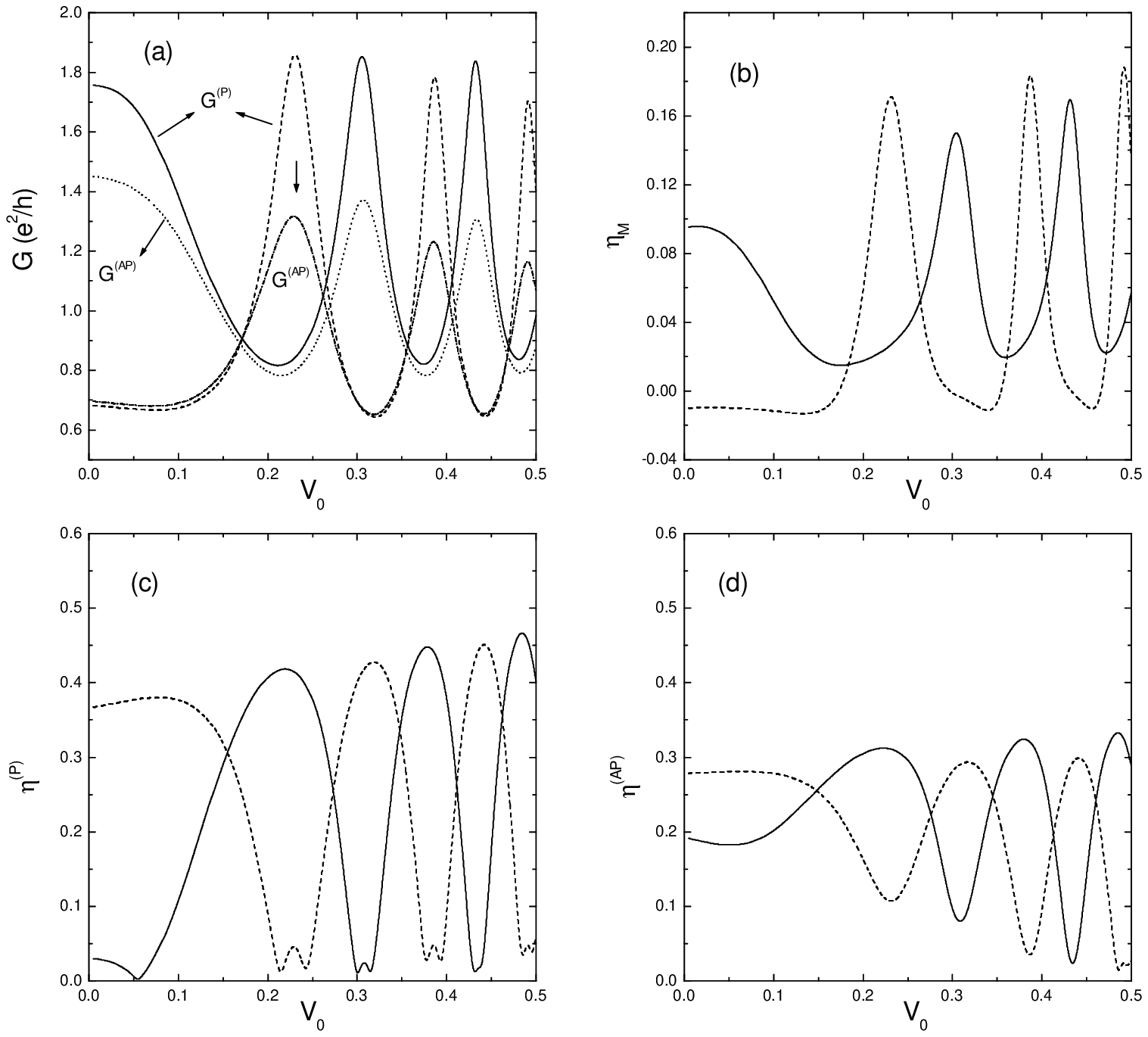}
\caption{
The changes of (a) the conductance $G^{(P)}$ and $G^{(AP)}$, (b) the magnetoconductance 
ratio $\protect\eta _{M}$, and (c) the spin injection coefficient 
$\protect\eta ^{(P)}$ and $\protect\eta ^{(AP)}$, 
with the variation of the bias voltage $V$ 
in two cases with different Rashba spin-orbit coupling constant. ( The
strength of Rashba spin-orbit coupling is characterized by the Rashba wave
number 
$k_{R}\equiv m_{s}\protect\alpha _{R}/\hbar ^{2}$. 
In Fig1.(a), $k_{R}=10^{7}$cm$^{-1}$ 
for the solid line and the dotted line, $k_{R}=5\times 10^{7}$cm$^{-1}$ 
for the dashed and the dash-dotted line. In
Fig.1(b)-(d), 
$k_{R}=10^{7}$cm$^{-1}$ 
for the solid line and 
$k_{R}=5\times10^{7}$cm$^{-1}$ 
for dashed line. Other parameters used were given in the text.
}
\end{figure}

Based on the formulas established above, in this section we will present
some numerical examples by considering some actual \ experimental
parameters. We will solve Eqs.(14)-(17) numerically by transfer matrix
method. In order to obtain the transfer matrix, it may be more convenient to
rewrite the wave function in the electrodes in a more general form as
following: 
\begin{eqnarray}
\Psi _{F}(x) &=&\sum_{\sigma =\pm }[a_{\sigma }^{(+)}\phi _{F,\sigma
,L}^{(+)}(x)|\sigma \rangle +a_{\sigma }^{(-)}\phi _{F,\sigma
,L}^{(-)}(x)|\sigma \rangle ],  \notag \\
\;x &<&0, \\
\Psi _{F}(x) &=&\sum_{\gamma =\pm }[b_{\gamma }^{(+)}\phi _{F,\gamma
,R}^{(+)}(x)|\gamma \rangle +b_{\gamma }^{(-)}\phi _{F,\gamma
,R}^{(-)}(x)|\gamma \rangle ],\;  \notag \\
x &>&L,
\end{eqnarray}%
If the spin of incident electron is $|\sigma \rangle $, one has $a_{\sigma
}^{(+)}=1$, $a_{\bar{\sigma}}^{(+)}=0$, $a_{\sigma }^{(-)}=r_{\sigma \sigma }
$, $a_{\bar{\sigma}}^{(-)}=r_{\sigma \bar{\sigma}}$, $b_{\gamma
}^{(+)}=t_{\sigma \gamma }$, $b_{\bar{\gamma}}^{(+)}=t_{\sigma \bar{\gamma}}$%
. $b_{\gamma }^{(-)}$ ( $\gamma =\pm $ ) will be set to be zero. From
Eq.(12) and Eqs.(20)-(21), at the interfaces between the F and S regions, $%
\Psi _{F}(x)$, $\Psi _{S}(x)$, $\hat{v}_{F}\Psi _{F}(x)$, and $\hat{v}%
_{S}\Psi _{S}(x)$ can be expressed as following: 
\begin{eqnarray}
\left[ 
\begin{array}{l}
\Psi _{F}(x)|_{x=0^{-}} \\ 
\hat{v}_{F}\Psi _{F}(x)|_{x=0^{-}}%
\end{array}%
\right]  &=&\hat{S}_{1}\left[ 
\begin{array}{l}
a_{+}^{(+)} \\ 
a_{-}^{(+)} \\ 
a_{+}^{(-)} \\ 
a_{-}^{(-)}%
\end{array}%
\right] ,  \notag \\
\left[ 
\begin{array}{l}
\Psi _{S}(x)|_{x=0^{+}} \\ 
\hat{v}_{S}\Psi _{S}(x)|_{x=0^{+}}%
\end{array}%
\right]  &=&\hat{S}_{2}\left[ 
\begin{array}{l}
c_{1,\beta } \\ 
c_{2,\beta } \\ 
c_{1,\bar{\beta}} \\ 
c_{2,\bar{\beta}}%
\end{array}%
\right] ,  \notag \\
\left[ 
\begin{array}{l}
\Psi _{S}(x)|_{x=L^{-}} \\ 
\hat{v}_{S}\Psi _{S}(x)|_{x=L^{-}}%
\end{array}%
\right]  &=&\hat{S}_{3}\left[ 
\begin{array}{l}
c_{1,\beta } \\ 
c_{2,\beta } \\ 
c_{1,\bar{\beta}} \\ 
c_{2,\bar{\beta}}%
\end{array}%
\right] ,  \notag \\
\left[ 
\begin{array}{l}
\Psi _{F}(x)_{x=L^{+}} \\ 
\hat{v}_{F}\Psi _{F}(x)|_{x=L^{+}}%
\end{array}%
\right]  &=&\hat{S}_{4}\left[ 
\begin{array}{l}
b_{+}^{(+)} \\ 
b_{-}^{(+)} \\ 
b_{+}^{(-)} \\ 
b_{-}^{(-)}%
\end{array}%
\right] ,
\end{eqnarray}%
where $\hat{S}_{i}$ $(i=1,2,3,4)$ are matrices, and the matrix elements of $%
\hat{S}_{i}$ can be written directly from Eqs.(12) and Eqs.(20)-(21). From
the matching condition (14)-(17) and Eq.(22), one gets that 
\begin{equation}
\left[ 
\begin{array}{l}
a_{+}^{(+)} \\ 
a_{-}^{(+)} \\ 
a_{+}^{(-)} \\ 
a_{-}^{(-)}%
\end{array}%
\right] =\hat{S}_{t}\left[ 
\begin{array}{l}
b_{+}^{(+)} \\ 
b_{-}^{(+)} \\ 
b_{+}^{(-)} \\ 
b_{-}^{(-)}%
\end{array}%
\right] ,
\end{equation}%
where $\hat{S}_{t}\equiv \hat{S}_{1}^{-1}\hat{S}_{2}\hat{S}_{3}^{-1}\hat{S}%
_{4}$ are the transfer matrix. Taking $b_{+}^{(-)}=0$ and $b_{+}^{(-)}=0$,
then from Eq.(23) one gets that 
\begin{equation}
\left[ 
\begin{array}{l}
b_{+}^{(+)} \\ 
b_{-}^{(+)}%
\end{array}%
\right] =\hat{T}\left[ 
\begin{array}{l}
a_{+}^{(+)} \\ 
a_{-}^{(+)}%
\end{array}%
\right] ,\hat{T}=\left[ 
\begin{array}{ll}
S_{t}(1,1) & S_{t}(1,2) \\ 
S_{t}(2,1) & S_{t}(2,2)%
\end{array}%
\right] ^{-1},
\end{equation}%
where $S_{t}(i,j)$ are the matrix elements of the transfer matrix $\hat{S}%
_{t}$. Since $a_{\sigma }^{(+)}=1$ and $a_{\bar{\sigma}}^{(+)}=0$ if the
spin of incident electron is $|\sigma \rangle $, then the transmission
coefficient can be got directly from Eq.(24) as following: $t_{++}=T(1,1)$, $%
t_{+-}=T(2,1)$, $t_{-+}=T(1,2)$, $t_{--}=T(2,2)$, where $T(i,j)$ are the
elements of the matrix $\hat{T}$. After the transmission coefficients are
obtained, the spin conductance of the device can be got from Eq.(18). In the
following we will focus on iron (Fe) as the ferromagnetic source and drain
and InAs as the semiconductor channel. In the ferromagnetic electrodes the
Fermi energy ( in the equilibrium state ) will be set to $E_{F}=2.469eV$ and
the exchange splitting energy be set to $\Delta =3.46eV$, appropriate for
Fe. The effective masses were set to $m_{f}=m_{e}$ ( for Fe ) and $%
m_{s}=0.036m_{e}$ ( for InAs ), and the band mismatch between the F and S
regions were set to $\delta E_{c}=2.0eV$. The length of the semiconductor
channel was set to be 1$\mu m$. The strength of Rashba spin-orbit coupling
will be characterized by a Rashba wave number $k_{R}\equiv m_{s}\alpha
_{R}/\hbar ^{2}$. For simplicity, we first assume that the interfacial
scattering potential is zero ( $\hat{U}=0$ ). In Figs.1(a)-(b) we have
plotted the changes of the total conductance $G^{(P)}$ and $G^{(AP)}$ and
the magnetoconductance ratio $\eta _{M}$ with the variation of the bias
voltage $V$ in two cases with different strengths of Rashba spin-orbit
coupling, and the changes of the spin injection coefficient $\eta ^{(P)}$
and $\eta ^{(AP)}$ with the variation of the bias voltage $V$ were also
plotted in Figs.1(c)-(d), respectively. From Figs.1(a)-(d) one can see that
in a large range of the bias voltage $V$, the conductance and the
magnetoconductance ratio and the spin injection coefficient all can be
changed significantly by tuning the Rashba spin-orbit coupling, i.e., the
structure described the Hamiltonian (1) may exhibit significant
spin-transistor effect in a large range of the bias voltage. But
Figs.1(a)-(d) show that the modulations of the conductance and the
magnetoconductance ratio and the spin injection coefficient due to Rashba
spin-orbit coupling may depend sensitively on the bias voltage, i.e., the
changes of the conductance and the magnetoconductance ratio and the spin
injection coefficient with the variation of Rashba spin-orbit coupling (
which can be tuned by changing the gate voltage ) may be very different
under different bias voltages. This can be seen more clearly from
Figs.2(a)-(c), where we have plotted the changes of the conductance $G^{(P)}$
and $G^{(AP)}$ and the magnetoconductance ratio $\eta _{M}$ and the spin
injection coefficient $\eta ^{(P)}$ with the variation of the Rashba
spin-orbit coupling constant ( characterized by the Rashba wave number $%
k_{R}\equiv m_{s}\alpha _{R}/\hbar ^{2}$ ) in two distinct cases with
different bias voltage $V$. From Figs.2(a)-(c) one can see clearly that the
bias voltage may have significant influence on the modulations of the
conductance and the magnetoconductance ratio and the spin injection
coefficient due to Rashba spin-orbit coupling. From theoretical viewpoints,
the spin-transistor effect due to Rashba spin-orbit coupling may depend
sensitively on the bias voltage because that the application of a finite
bias voltage will not only change the energies of incident electrons ( as in
usual electronic devices ) but also have influence on the gate-voltage
controlled spin precession in the S channel of the device. The reason for
this is that when a finite bias voltage is applied between two terminals of
a spin FET, a longitudinal electric field will be established in the
semiconductor channel, and due to the presence of this longitudinal electric
field, spin precessions of injected spin-polarized electrons in the S
channel will depend not only on the gate-voltage controlled Rashba
spin-orbit coupling but also depend on the bias voltage. This can be seen
clearly from the formulas presented in section II, where we have shown that
in the presence of a finite bias voltage, the spinor wave function in the S
region will depend not only on the Rashba spin-orbit coupling constant but
also depend on the bias voltage. So, in order to describe correctly the spin
precession states of injected spin-polarized electrons in the semiconductor
channel of a spin FET, the interplay of the gate-voltage controlled Rashba
spin-orbit coupling and the longitudinal electric field induced by the
application of a finite bias voltage should be described in a unified way,
as was shown in section II. Next, we consider the effect of interfacial
scattering. The strength of interfacial scattering can be characterized by
two dimensionless parameters defined by $z_{1}=(U_{1}/\hbar )\sqrt{%
2m_{f}/E_{F}}$ and $z_{2}=(U_{2}/\hbar )\sqrt{2m_{f}/E_{F}}$, where $U_{1}$
and $U_{2}$ are the diagonal and off-diagonal elements of the interfacial
scattering potential matrix $\hat{U}$. The parameters $z_{1}$ and $z_{2}$
represent the strengths of spin-conserving and spin-flip interfacial
scattering, respectively. The effect of interfacial scattering can be seen
from Figs3.(a)-(c), where we have plotted the changes of the conductance $%
G^{(P)}$ and the spin injection coefficient $\eta ^{(P)}$ and $\eta ^{(AP)}$
with the variation of the bias voltage $V$ in the presence of (
spin-conserving and/or spin-flip ) interfacial scattering. Fig.3(a) shows
that both spin-conserving and spin-flip interfacial scatterings will
decrease substantially the conductance of the device, except at some special
values of the bias voltage. Figs.3(b)-(c) show that interfacial scattering
may decrease the spin injection coefficient if the two ferromagnetic
contacts of the device are in parallel configuration and will enhance the
spin injection efficiency if the two ferromagnetic contacts of the device
are in antiparallel configuration. An interesting fact that can be seen from
Figs.3(b)-(c) is that the enhancement of the spin injection efficiency due
to interfacial scattering may be very substantial when the two ferromagnetic
contacts of the device are in antiparallel configuration, compared with the
decrease of the spin injection efficiency due to interfacial scattering when
the contacts of the device are in parallel configuration.

\begin{figure}[tbp]
\includegraphics*[width=6.7cm]{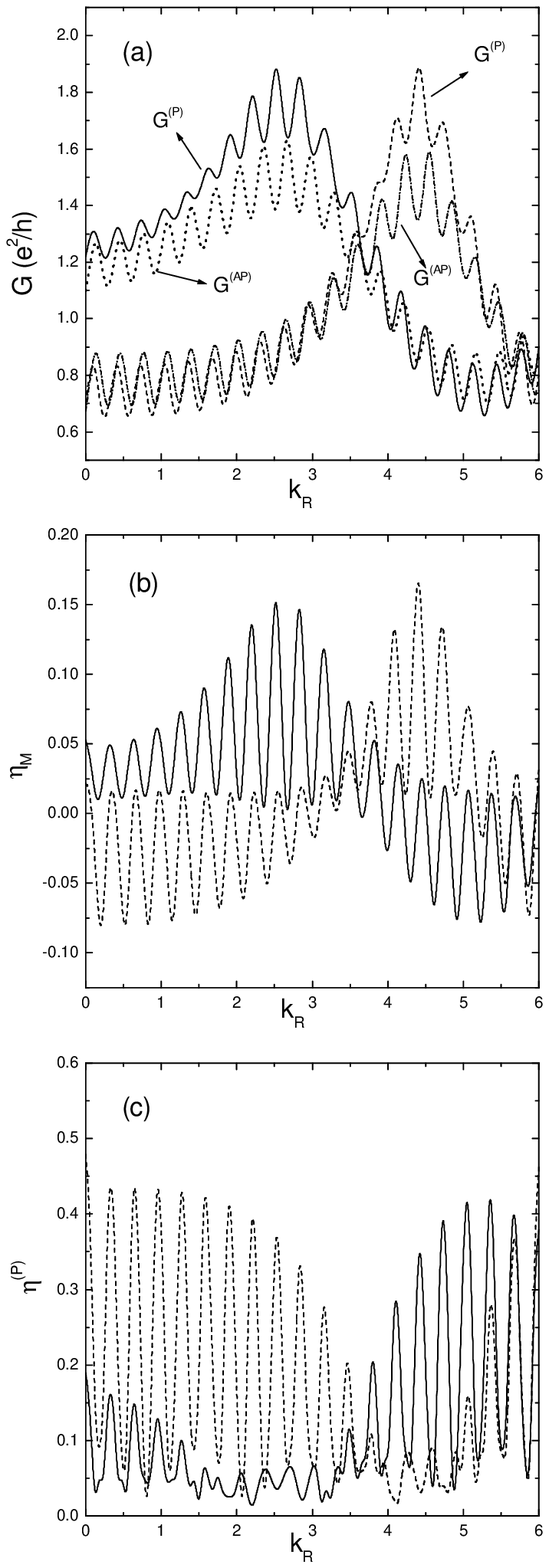}
\caption{
The changes of (a) the
conductance $G^{(P)}$ and $G^{(AP)}$, (b) the magnetoconductance ratio $%
\protect\eta _{M}$, and (c) the spin injection coefficient $\protect\eta %
^{(P)}$, with the variations of the Rashba wave number $k_{R}$ in two cases
with different bias voltages. ( In Fig2.(a), $V=0.1$V for the solid and the
dotted lines, $V=0.2$V for the dashed and the dash-dotted lines. In
Fig.2(b)-(c), $V=0.1$V for the solid and $V=0.2$V for the dashed line. Other
parameters used were given in the text. The changes of the spin injection
coefficient $\protect\eta ^{(AP)}$ with the variations of $k_{R}$ is similar
as was shown in Fig.(c) and were not plotted. 
}
\end{figure}

\begin{figure}[tbp]
\includegraphics*[width=6.7cm]{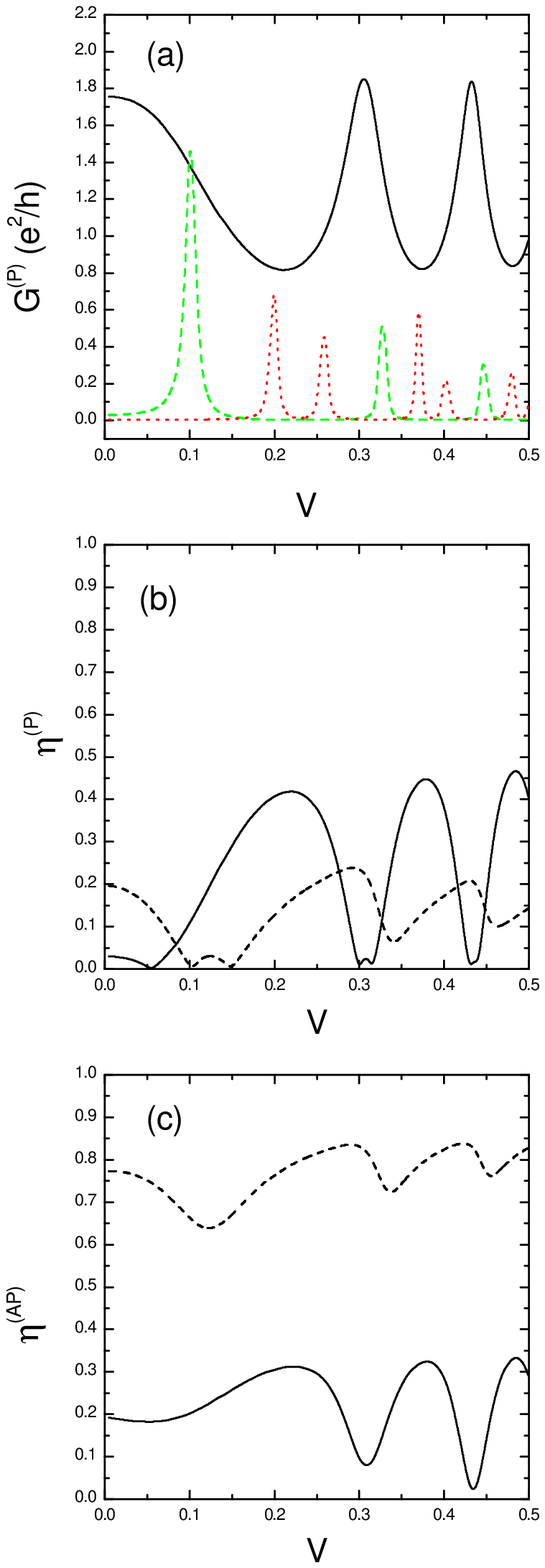}
\caption{
The changes of the conductance $%
G^{(P)}$and the spin injection coefficient $\protect\eta ^{(P)}$ and $%
\protect\eta ^{(AP)}$ with the variations of the bias voltage $V$ in the
presence of interfacial scattering. ( The strength of interfacial scattering
are characterized by two dimensionless parameters defined by $z_{1}\equiv
(U_{1}/\hbar )\protect\sqrt{2m_{f}/E_{F}}$ and $z_{2}\equiv (U_{2}/\hbar )%
\protect\sqrt{2m_{f}/E_{F}}$. In Fig.3(a), $z_{1}=0$ and $z_{2}=0$ for the
solid line; $z_{1}=10$ and $z_{1}=0$ for the dotted line; $z_{1}=0$ and $%
z_{2}=10$ for the dashed line. In Fig3.(b)-(c), $z_{1}=0$ and $z_{2}=0$ for
the solid line; $z_{1}=2$ and $z_{2}=5$ for the dashed line. Other
parameters used were given in the text. The changes of the conductance $%
G^{(AP)}$ with the variations of the bias voltage $V$ is similar as was
shown in Fig.1(a) and were not plotted. 
}
\end{figure}

In conclusion, in this paper we have discussed the influence of finite bias
voltages on the conductance modulations in spin FETs due to Rashba
spin-orbit coupling. We have shown that when a finite bias voltage is
applied between both terminals of a spin FET, the conductance modulation in
the device due to Rashba spin-orbit coupling may depend sensitively on the
bias voltage, and in order to describe correctly the spin precession states
of injected spin-polarized electrons in the semiconductor channel of the
device, the interplay of the gate-voltage controlled Rashba spin-orbit
coupling and the bias voltage should be described in a unified way.

This work was supported by a grant from the Research Grant Council of Hong
Kong, China.

\bigskip 

\end{document}